\documentclass{aa}
\usepackage{graphicx}

\def\lsim{\stackrel{<}{\scriptstyle \sim}}

\begin{document}

%\thesaurus {06 (04.19.11, 08.02.2, 08.02.4, 08.05.1, 08.06.3)}

\title{Asteroids as radial velocity and resolving power standards for medium and high resolution spectroscopy
}

\authorrunning{T. Zwitter et al.}
\titlerunning{Asteroids as radial velocity and resolving power standards }

\author{
       Toma\v{z} Zwitter\inst{1}
\and   Francois Mignard\inst{2}
\and   Francoise Crifo\inst{3}
       }
\offprints{T. Zwitter}

\institute {
University of Ljubljana, Department of Physics, Jadranska 19, 1000 Ljubljana, Slovenia 
\and
Observatoire de la C\^{o}te d'Azur, Cassiop\'{e}e, CNRS UMR 5202, BP 4229, 06304 Nice Cedex 4, France
\and 
Observatoire de Paris, GEPI, 5 Place Jules Janssen, 92195 Meudon, France
}
\date{Received date ...... / accepted date .......}

\abstract{Echelle spectra of 10 bright asteroids are presented and compared against an 
observed twilight spectrum and a computed Solar spectrum. Spectra covering a 2130~\AA\ 
spectral range centered on $\lambda = 5785$~\AA\ are of high resolving power and high 
signal to noise ratio. So we focus on a comparison of detailed properties of spectral 
lines and not on albedo variations. It is shown that the normalized Solar and 
asteroid spectra are identical  except for radial velocity (RV) shifts 
which  can be predicted at the accuracy level of 1~m~s$^{-1}$. 
So asteroids are proposed as new and extremely accurate radial velocity 
standards.  Predicted and measured RVs of observed asteroids indeed match 
within limits of accuracy of the instrument. There are numerous absorption 
lines in the reflected Solar spectrum. This allows a direct mapping of the resolving power 
of a spectrograph between and along echelle spectral orders.  So asteroid spectra
can be used to test wavelength calibration and resolving power  of spectrographs 
on the ground as well as in space, including the Gaia mission of ESA. All 
spectra are also given in an electronic form.

}
\maketitle

\section{Introduction}

Spectroscopic observations are usually wavelength calibrated using spectra 
of standard calibration lamps. Still, it is desirable to check this 
calibration with real objects on the sky in order to discover possible 
instrumental effects. These effects include: the calibration and the stellar 
beams not being exactly parallel when they enter the spectrograph slit; 
uneven illumination of the slit, which causes spectral shifts and therefore 
errors in the measurement of radial velocity; flexures and/or temperature 
instabilities (present in most of the spectrographs which are mounted 
directly on the telescope), which could cause significant variation of velocity 
and/or spectral shape if not monitored and properly accounted for. 

IAU Commission 30 "Radial velocities" has defined a list of standard stars
with accurate velocities  constant over many years. The list of 
Elodie-CORAVEL high-precision standard stars   (published on
obswww.unige.ch/$\sim$udry/std/std.html) contains a total of 140 stars, 38 of them
with a  radial velocity error $\lsim 50$~m~s$^{-1}$\ (Udry, Mayor \&\ Queloz
1999). Most of these stars are bright ($V < 7$) so they may be unsuitable 
as radial velocity standards for spectrographs on  large telescopes due to 
saturation problems. 
The same is true for the now quite numerous stars which are being searched 
for extrasolar planets. Also, the RV residuals for some of these stars show a 
peculiar behaviour possibly due to yet undiscovered planetary companions, so 
their RV is not predictable accurately enough for calibration purposes. What 
is more, in some cases only variation but not the absolute value for RV has 
been derived and published.
Nordstrom et al.\ (2004) published radial velocities of the first 
15\%\ of stars in the Hipparcos catalogue, part of which may be used as references 
if an error up to 300~m~s$^{-1}$\ is acceptable. The sample
which is being observed by the RAVE survey (Steinmetz et al.\ 2006) is much larger 
and fainter, but of moderate accuracy. Most of the stars are observed only once, 
so some of the radial velocities may belong to (the yet undiscovered) binary 
stars. 

Here we propose to use asteroids as objects for which radial velocity at the 
time of observation can be easily calculated to an extreme accuracy level. 
To the best of our knowledge no high resolution spectra of even the brightest 
asteroids exist in the literature. So we followed our initial suggestion 
(Zwitter \&\ Crifo 2003, Mignard 2003) and obtained high resolution and high signal to 
noise echelle spectra of 10 bright asteroids. Measurements of their radial 
velocities are compared to the accurate calculated values. Normalized asteroid 
spectra are also compared against both theoretical and observed (twilight) 
Solar spectra for each of the echelle orders. We show that asteroid spectra 
have identical spectral lines to twilight spectra. Excellent distribution of Solar 
spectral lines over the whole UV, optical and IR domain allows for a more accurate 
assessment of wavelength accuracy and resolving power than the conventional 
methods which use telluric lines. 

The next section describes the observations and data reduction procedures. 
In section 3 we briefly describe the computations necessary to derive an 
accurate prediction of asteroid radial velocity at the time of observation. 
Observed asteroid and Solar spectra are compared in Sections 4 and 5, 
to be followed by Conclusions.

\begin{table}[!Ht]
\tabcolsep 0.08truecm
\label{table1}
\caption{Wavelength ranges of individual spectral interference orders of the 
Asiago echelle spectrograph.}
\begin{tabular}{clclcclcl} \hline \hline
&&&&&&&&\\
order&\multicolumn{3}{c}{range [\AA]}&\ \ \ \ &order&\multicolumn{3}{c}{range [\AA]}\\
&&&&&&&&\\
\hline
&&&&&&&&\\ 
33&6720   &--& 6850&&
41&5410   &--& 5510\\
34&6520   &--& 6650&&
42&5280   &--& 5380\\
35&6330   &--& 6460&&
43&5150   &--& 5250\\
36&6160   &--& 6270&&
44&5040   &--& 5130\\
37&5990   &--& 6110&&
45&4930   &--& 5020\\
38&5830   &--& 5940&&
46&4820   &--& 4910\\
39&5680   &--& 5790&&
47&4720   &--& 4810\\
40&5540   &--& 5650&&
  &       &  &     \\
  &       &  &     &&
  &       &  &     \\
 \hline
\end{tabular}
\end{table}

\begin{table}[!Ht]
\tabcolsep 0.08truecm
\label{table2}
\caption{Log of asteroid observations. $N$ is the number of exposures, 
$t$ is the total observing time, {\it Date} is the truncated 
date of observation and $V$ is the apparent V magnitude (derived from
the Lowell Observatory  asteroid database). 
$S/N$ is the signal to noise ratio per 
0.128~\AA\ wavelength bin at  $6070\pm 4$~\AA. }
%V magnitudes from http://asteroid.lowell.edu/cgi-bin/koehn/asteph
\begin{tabular}{rllrccrr} \hline \hline
&&&&&&&\\
\multicolumn{2}{l}{Name}&$N$&$t$~[s]&\ \ &Date&V& $S/N$\\ 
&&&&&&&\\ \hline
&&&&&&&\\ 
1 & Ceres   & 3 & 3000 && 2005 04 03.12 & 7.6& 188 \\
2 & Pallas  & 4 & 3300 && 2005 04 02.95 & 7.3& 242 \\
3 & Juno    & 3 & 3000 && 2004 09 25.81 &10.4&  81 \\
4 & Vesta   & 4 & 3200 && 2004 09 26.00 & 6.3& 204 \\
9 & Metis   & 3 & 3000 && 2004 09 25.45 & 9.2& 153 \\
21& Lutetia & 3 & 3600 && 2004 09 26.13 &10.7&  98 \\
27& Euterpe & 3 & 3600 && 2004 09 27.13 &10.2& 214 \\
40& Harmonia& 3 & 3600 && 2004 09 25.09 & 9.8&  94 \\
49& Pales   & 3 & 3600 && 2004 09 29.17 &11.5&  71 \\
80& Sappho  & 3 & 2700 && 2004 09 28.64 &11.1&  84 \\
  & dawn sky& 3 &  480 && 2004 09 26.20 &    & 340 \\
  &         &   &      &&               &    &     \\
\hline
\end{tabular}
\end{table}

\begin{table*}[!Ht]
\tabcolsep 0.08truecm
\label{table3}
\caption{Radial velocities measured from individual spectral orders. Asteroid's radial velocity 
was determined as a mean of results for orders 33--43, skipping the 2 most deviant values (see text). 
}
\begin{tabular}{rlrrrrrrrrrrrrrrr} \hline \hline
&&&&&&&&&&&&&&&&\\
\multicolumn{2}{c}{Name}&\multicolumn{15}{c}{radial velocity from echelle order}\\
 &          &\multicolumn{1}{c}{33}&\multicolumn{1}{c}{34}&\multicolumn{1}{c}{35}&\multicolumn{1}{c}{36}&
	     \multicolumn{1}{c}{37}&\multicolumn{1}{c}{38}&\multicolumn{1}{c}{39}&\multicolumn{1}{c}{40}&
	     \multicolumn{1}{c}{41}&\multicolumn{1}{c}{42}&\multicolumn{1}{c}{43}&\multicolumn{1}{c}{44}&
	     \multicolumn{1}{c}{45}&\multicolumn{1}{c}{46}&\multicolumn{1}{c}{47}\\
\hline
&&&&&&&&&&&&&&&&\\
1 &Ceres     &-12.03  &-12.31  &-12.29  &-12.36  &-12.46  &-12.42  &-12.62  &-12.49  &-13.18  &-13.45  &-13.72  &-13.32  &-13.47  &-13.73  &-14.00  \\%01270129  
2 &Pallas    & 14.14  & 14.18  & 13.96  & 13.78  & 13.47  & 13.71  & 13.46  & 13.68  & 13.04  & 13.19  & 12.82  & 13.64  & 13.64  & 13.67  & 13.18  \\%01110114 
3 &Juno      & 15.34  & 14.64  & 15.13  & 15.20  & 15.16  & 15.66  & 15.30  & 15.36  & 15.17  & 15.16  & 15.15  & 15.82  & 15.88  & 16.41  & 16.17  \\%00370039 
4 &Vesta     &  9.73  &  9.69  &  9.95  &  9.89  &  9.82  & 10.14  & 10.28  & 10.56  & 10.23  & 10.24  & 10.36  & 10.84  & 10.79  & 11.12  & 11.19  \\%00570060 
9 &Metis     &  0.73  &  0.44  &  0.80  &  0.57  &  0.54  &  0.71  &  0.76  &  0.97  &  0.64  &  0.76  &  0.87  &  1.44  &  1.56  &  1.95  &  1.77  \\%00520054  
21&Lutetia   & -8.24  & -9.15  & -8.66  & -8.86  & -9.09  & -8.26  & -8.63  & -8.79  & -8.67  & -8.60  & -8.70  & -8.11  & -8.27  & -7.90  & -7.98  \\%00680070 
27&Euterpe   &-19.54  &-19.80  &-19.69  &-19.78  &-19.93  &-19.38  &-19.55  &-19.47  &-19.77  &-19.59  &-19.68  &-18.95  &-18.78  &-18.51  &-18.65  \\%01170119 
40&Harmonia  & -6.50  & -6.59  & -6.51  & -6.73  & -6.87  & -6.74  & -6.69  & -6.46  & -6.76  & -6.60  & -6.75  & -6.16  & -6.05  & -5.73  & -5.55  \\%00250027 
49&Pales     &-13.87  &-13.34  &-13.43  &-13.76  &-13.81  &-12.92  &-13.67  &-13.52  &-13.86  &-13.53  &-13.57  &-13.13  &-12.66  &-12.52  &-12.78  \\%01880190 
80&Sappho    &-11.01  &-10.54  &-10.46  &-10.76  &-10.76  &-10.18  &-10.48  &-10.33  &-10.73  &-10.59  &-10.57  &-10.04  & -9.78  & -9.48  & -9.61  \\%01500152 
  &twilight  & -0.12  & -0.43  & -0.35  & -0.52  & -0.66  & -0.33  & -0.51  & -0.34  & -0.83  & -0.76  & -0.86  & -0.34  & -0.28  & -0.09  & -0.21  \\%00820084 
&&&&&&&&&&&&&&&&\\ \hline
\end{tabular}
\end{table*}

\begin{table*}[!Ht]
\tabcolsep 0.08truecm
\label{table4}
\caption{Calculated radial velocities compared to observations. Columns (3) and (4) give 
Sun--asteroid and asteroid to Asiago observatory distances in a.u. The next columns are  
velocities in km~s$^{-1}$: column (5) is radial velocity of asteroid vs.\ Sun, column (6) is radial 
velocity of asteroid vs. Asiago observatory and column (7) is 
their sum. Column (8) gives radial velocity as derived from observations (see Table 3).
}
\begin{tabular}{rllllrrrrr} \hline \hline
&&&&&&&&&\\
\multicolumn{2}{c}{Name}&\multicolumn{1}{c}{JD(UTC)}&\multicolumn{1}{c}{d$_\odot$}&
\multicolumn{1}{c}{d$_{tel.}$}&\multicolumn{1}{c}{RV$_\odot$}&\multicolumn{1}{c}{RV$_{tel}$}&
\multicolumn{1}{c}{RV(calc)}&\multicolumn{1}{c}{RV(obs)}&\multicolumn{1}{c}{O$-$C}\\
\multicolumn{2}{c}{(1)}&\multicolumn{1}{c}{(2)}&\multicolumn{1}{c}{(3)}&\multicolumn{1}{c}{(4)}&
\multicolumn{1}{c}{(5)}&\multicolumn{1}{c}{(6)}&\multicolumn{1}{c}{(7)}&\multicolumn{1}{c}{(8)}&
\multicolumn{1}{c}{(9)}\\
\hline
&&&&&&&&&\\
 1&Ceres   & 2453463.619167&2.66095& 1.83302&  1.3092&--13.9723&--12.6631&     --12.43 & 	0.23\\ %01270129
 2&Pallas  & 2453463.448085&2.38562& 1.41984&  3.8052&   9.6790&  13.4842&	 13.80 & 	0.32\\ %01110114
 3&Juno    & 2453274.309769&2.82011& 2.50488&--4.3485&  19.1563&  14.8078&	 15.20 & 	0.39\\ %00370039
 4&Vesta   & 2453274.498461&2.38032& 1.41049&  1.7008&   8.0025&   9.7033&	 10.02 & 	0.32\\ %00570060
 9&Metis   & 2453274.449896&2.30037& 1.32308&--2.3264&   3.2949&   0.9685&	  0.70 &      --0.27\\ %00520054
21&Lutetia & 2453274.628148&2.17825& 1.34490&  2.6739&--10.9357&  -8.2618&	--8.68 &      --0.42\\ %00680070
27&Euterpe & 2453275.632188&2.10125& 1.28461&--2.9481&--16.5471&--19.4952&     --19.65 &      --0.15\\ %01170119
40&Harmonia& 2453273.592419&2.16246& 1.20745&  0.1823&  -6.9727&  -6.7904&	--6.66 & 	0.13\\ %00250027
49&Pales   & 2453277.665752&2.36502& 1.52997&--0.3002&--13.6541&--13.9543&     --13.60 & 	0.35\\ %01880190
80&Sappho  & 2453276.636134&1.92316& 1.26508&  2.7895&--13.4288&--10.6393&     --10.56 & 	0.08\\ %01500152
&&&&&&&&&\\ \hline
\end{tabular}
\end{table*}

\section{Observations and Data Reduction}

Spectroscopic observations have been collected with the echelle$+$CCD 
spectrograph of the 1.82~m telescope operated by Osservatorio Astronomico 
di Padova atop of Mt.\ Ekar (Asiago, Italy). The spectrograph is directly 
mounted on the Cassegrain F/9 focus of the telescope. The light enters 
the spectrograph through a 150~$\mu$m (1.9 arcsec) slit which is normally 
mounted along the P.A.=90$^\mathrm{o}$ or 270$^\mathrm{o}$ angle, i.e. in the 
E-W direction. The slit length of 12.6 arcsec allows the asteroid to be placed 
at one side and to collect useful sky background at the other side of the slit 
length. The position 
of the object on the slit is monitored by a red sensitive TV guider. CCD detector 
is a thinned broad-band coated E2V CCD47-10 with 1k$\times$1k square pixels of 
13~$\mu$m and with quantum efficiency $>70$\%\ in the interval of  3800 to 7500~\AA. 
Here we use the 15 echelle orders which cover the  4720 -- 6850~\AA\  range with small 
gaps between the orders (see Table~1). Redder orders are plagued by telluric absorptions of the 
Earth atmosphere, while those to the blue have difficulties with accurate determination 
of the continuum level. CCD is cooled with liquid nitrogen which renders the dark 
current negligible. The readout noise is $\sim 9$ electrons. Spectra are 
wavelength calibrated from spectra of a ThAr lamp which uniformly illuminates 
the slit. The resolving power of the spectrograph is  $\sim 24,000$  
at central wavelengths of echelle orders, with notable degradation towards the edges 
(see below) due to optical distortions. Wavelength sampling is from 0.0955~\AA /pix 
in spectral order 47, to 0.1357~\AA /pix in the order 33. So the resolution element equals 
2.0 pixels.

Observations of 10 bright asteroids were obtained. Table~2 summarizes the 
observing log. To observe a representative 
Solar spectrum integrated over the Solar disk we obtained a series of dawn 
twilight sky exposures. The telescope guiding was switched off during twilight exposures 
in order to minimize spectrograph flexures and light from chance superposition 
stars entering the slit. Each asteroid was observed 
in a series of consecutive exposures with a ThAr calibration lamp spectra 
obtained at the start and end of the series. Each spectrum in a series was 
independently reduced using standard IRAF echelle routines. The routine {\it apall }
was used to trace and extract the spectrum and to subtract the sky background which was 
assumed to be the median of pixels at the same wavelength and at the part of the 
slit not illuminated by the object. The wavelength calibration was done using both images 
of calibration lamp spectra. Lamp spectra were extracted with 
the aperture and tracing information from the object exposure. Wavelength 
solutions of both lamps were combined using an average weighted by the 
difference in time at mid exposure between the object and each of the lamp 
spectra. A few hot stars with very high rotational velocity were observed 
during each of the observing runs. Among them we used for the final reductions  
$\lambda$~Eri spectra for the September 2004 observing run and HD~149757 for 
the April 2005 run. A high order cubic spline fit was used to 
remove stellar features and thus obtain a normalized telluric spectrum. 
Each order of observed spectra of asteroids and twilight were normalized using a 5-piece cubic 
spline fit with 2$\sigma$ (lower) and 3$\sigma$ (upper) rejection limits and with 
10 iterations using a growing radius of 1~pixel. The normalized spectra were median 
combined and the normalized telluric spectrum was divided out to finally obtain the 
normalized spectrum of a given asteroid or twilight. One of the principal goals of our analysis 
is to compare a high resolution twilight spectrum to those of asteroids. We note that 
intensities of individual telluric features can be variable in time, so any spectral 
range with normalized telluric spectral intensities outside the 0.99 -- 1.01 range was 
excluded from further analysis. This criterion rejected 488 wavelength bins, 
so that the final spectra contain normalized intensities at 13,665 wavelengths.   

Radial velocities were measured against a theoretical spectrum which was calculated using the 
Kurucz models. We used the latest version of the model of the Solar atmosphere which was kindly 
provided to us by F.\ Castelli. The model was computed with the Atlas~9 code using the chemical 
abundances from Grevesse \&\ Sauval (1998). Opacity distribution functions were recalculated for 
these specific abundances (Castelli \&\ Kurucz 2004). The effective temperature is 5777~K, 
$\log g = 4.4377$, mixing length scale height is 1.25 and rotational velocity 
is 2.0~km~s$^{-1}$. The overshooting option was not used. The spectrum was calculated with the Linux 
implementation of the SYNTHE code (Sbordone et al.\ 2004). Normalized theoretical spectrum calculated 
at $R=500,000$ and degraded to the resolving power of 24,000 is reported in column 4 of Table~5.

Measurement of radial velocity of a given asteroid was done for each spectral order separately using 
the RVSAO package (Kurtz et al.\ 1992). The values of radial velocities were consistent between the orders, 
but the bluest 4 orders 
showed a distinct trend (see below). So the mean of observed velocities in orders 33--43 (with values of 
two most deviant orders not taken into account) was used as a final observed radial velocity of an asteroid.

\section{Calculation of radial velocity}

Radial velocity of an asteroid at a given moment can be calculated 
to extreme accuracy. This is because most asteroids 
have been observed during many ($\sim 30$) oppositions so that 
their positions are known to within 1 arcsec for several years in 
advance. Assuming a typical space motion of $\sim 30$~arcsec~h$^{-1}$
this corresponds to a time error of only 120~s. The radial velocity 
of an asteroid changes by $\approx 30$~km~s$^{-1}$ in 3 months or 
0.5~m~s$^{-1}$ in 120~s. So we may conclude that radial velocity can 
be calculated at an 1~m~s$^{-1}$ level of accuracy. 

Equatorial rotational velocity of an asteroid amounts to 22~m~s$^{-1}$ 
for an asteroid with a diameter of 250~km and with a 10~hour rotational 
period. For the largest and brightest asteroids it is somewhat larger, reaching a
maximum of 84 m~s$^{-1}$\ for Ceres.
Most large asteroids are nearly spherical in shape. Also reflectance 
for most of them (with an exception of Vesta) is quite constant over the whole surface. So 
the effect of rotation would be a very moderate broadening of the 
reflected Solar lines. This effect would be very difficult to measure 
because of the much larger temperature and rotational broadening of intrinsic 
Solar lines. 
However during the GAIA mission, which is the main aim of this work,
main belt asteroids will not be observed at opposition, but at a 
phase angle $\sim 20^\mathrm{o}$, and always less than 30$^\mathrm{o}$. Gibbous 
shape of the illuminated surface of such asteroids means that a crescent-shaped 
part near the limb with possibly high radial velocity is not illuminated. 
This introduces a small wavelength shift of the integrated reflected spectrum 
if an asteroid rotates about an axis pointing away from Earth.  
Assuming a spherical asteroid with rotation axis in the plane
of the sky, the surface of this missing crescent projected on the plane
of the sky may reach up to 14\%\ of the illuminated face at a 30$^\mathrm{o}$ 
phase angle (only 6\%\ at 20$^\mathrm{o}$). As the rotational velocity is always
smaller than 100 m~s$^{-1}$\ (see above), this results in a change in the line profile and
a Doppler shift smaller than 15 m~s$^{-1}$. Notice also that other physical
parameters, such as irregularities in shape and reflectance, would also
manifest in radial velocity shift. A hypothetical cigar-shaped asteroid with reflective 
and black surfaces on the opposite sides from the center of gravity would 
cause a shift with an amplitude of 1.3~m~s$^{-1}$ if its longer axis equals 
12~km, the rotation period is 4~hours and the rotation axis is in the plane of the sky.
Note that rapid rotation of such asteroid would be known due to photometric 
variability. Accurate treatment of asteroid rotation is 
beyond the scope of this paper. We note however that small radial 
velocity shifts caused by rotation can be in principle used to disentangle 
degeneracy between shape irregularity and reflectance variability in interpretation 
of asteroid light curves.

Calculation of a predicted radial velocity of an asteroid at the time 
of observation includes the following steps: 
\begin{itemize}
\item[(a)] A photon reaching the Earth at time $t_2$ has left the asteroid at the
retarded time $t_1 = t_2- r/c$, where $r$ is the distance between the
asteroid at the time of photon emission and the observer at its reception. The retarded position is computed by a
straightforward iteration of the asteroid orbital motion. The Doppler velocity 
$\mathrm{RV}_\mathrm{tel}$ results from the combination of the geocentric radial velocity of the 
asteroid at time $t_1$ and that of the observer due to the rotation of the Earth 
at time $t_2$.
\item[(b)] Radial velocity $\mathrm{RV}_\odot$ of an 
asteroid to the Sun is determined from its orbit at time $t_1$. 
\item[(c)] The total radial velocity shift of the Solar spectrum reflected 
by the asteroid and observed by the Earth observer is, for small velocities,
$\mathrm{RV} = \mathrm{RV}_\odot + \mathrm{RV}_\mathrm{tel}$.
\end{itemize}
We estimate that the values of calculated radial velocities as given in 
Table~4 are accurate to within 1~m~s$^{-1}$.
The radial velocity of the Sun which was observed in the scattered light 
of the dawn sky was calculated similarly. We note that the exact position 
of scatter is unknown. But since it lies within the 10~km thick layer 
of Earth atmosphere this uncertainty is negligible.

\begin{figure}[t]
%\centerline{\includegraphics[angle=270,width=9.2cm]{figures/fig_ordererrors.eps}}
\centerline{\includegraphics[angle=270,width=9.2cm]{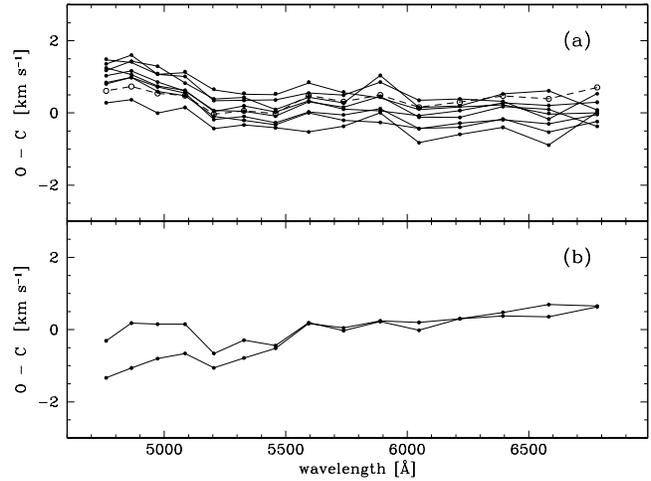}}
\label{fig1}
\caption[]{Difference between observed velocity for the given echelle order
and the calculated values. Top panel is for September 2004 observations, the bottom 
for the April 2005 ones. The twilight spectrum at dawn (dashed line) was observed 
only in the September 2004 run.
}
\end{figure}

\begin{figure}
%\centerline{\includegraphics[angle=270,width=9.2cm]{figures/plotkumul.eps}}
\centerline{\includegraphics[angle=270,width=9.2cm]{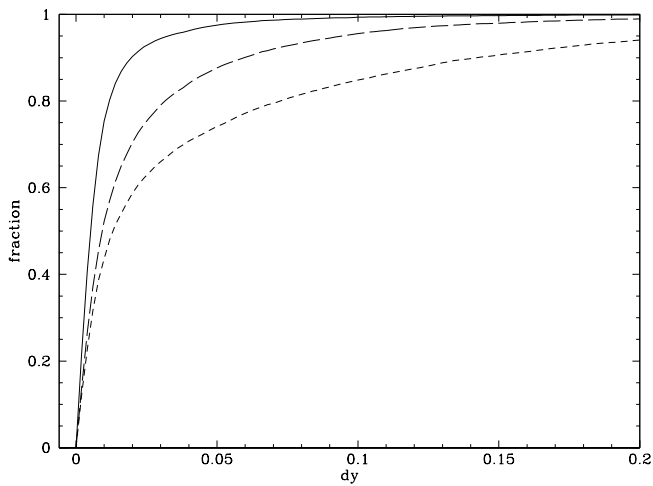}}
\label{fig2}
\caption[]{Cumulative plot of a fraction of wavelength points with two normalized 
spectra differing by less than a given amount $dy = abs(I_1-I_2)/I_2$. The pairs are
mean asteroid and twilight spectrum (solid line), theoretical Kurucz and twilight 
spectrum (long dashes), and intensities of twilight spectrum compared to a unit 
intensity (short dashes). 
}
\end{figure}

\begin{figure}[t]
%\centerline{\includegraphics[angle=270,width=9.2cm]{figures/respowerfigure01170119.eps}}
\centerline{\includegraphics[angle=270,width=9.2cm]{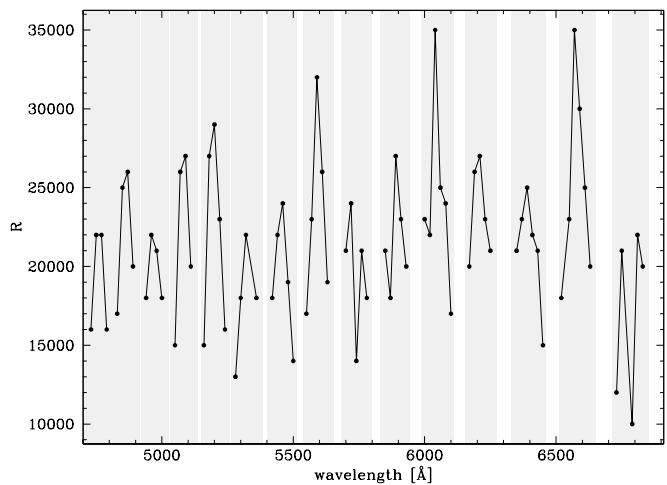}}
\label{fig3}
\caption[]{Resolving power $R \equiv \lambda / d\lambda$ of the spectrum of 27 Lutetia as a function of wavelength.
Points present $R$ of the theoretical Solar spectrum which gave the best match to the 
observed spectrum in the interval $\pm 10$~\AA\ from the point. Points within the given 
echelle spectral order are connected and the area of the order is shaded.
}
\end{figure}

\begin{table*}[!Ht]
\tabcolsep 0.08truecm
\label{table5}
\caption{  Observed spectra of twilight and asteroids rebinned to the same 
wavelength bins, continuum normalized and Doppler shifted to zero radial 
velocity.
Columns give wavelength (1), twilight (2) and median asteroid intensities (3), normalized flux from the 
Kurucz Solar model (4), and intensities of individual asteroid spectra (5-14). 
Complete table is available in electronic form through CDS.
}
\begin{tabular}{cccccccccccccc} \hline \hline 
&&&&&&&&&&&&\\
$\lambda$ [\AA]&twilight&asteroids&Kurucz&1\,Ceres& 2\,Pallas& 3\,Juno& 4\,Vesta& 9\,Metis & 21\,Lutetia & 27\,Euterpe & 40\,Harmonia & 49\,Pales & 80\,Sappho\\
\multicolumn{1}{c}{(1)}&\multicolumn{1}{c}{(2)}&\multicolumn{1}{c}{(3)}& \multicolumn{1}{c}{(4)}& \multicolumn{1}{c}{(5)}& 
\multicolumn{1}{c}{(6)}& \multicolumn{1}{c}{(7)}& \multicolumn{1}{c}{(8)}& \multicolumn{1}{c}{(9)}& \multicolumn{1}{c}{(10)} & 
\multicolumn{1}{c}{(11)} & \multicolumn{1}{c}{(12)} & \multicolumn{1}{c}{(13)} & \multicolumn{1}{c}{(14)}\\	       
\hline
&&&&&&&&&&&&\\
6024.0667 &0.731 &0.686 &0.716 &0.712 &0.729 &0.711 &0.683 &0.620 &0.689 &0.672 &0.688 &0.670 &0.673\\
6024.1951 &0.821 &0.797 &0.839 &0.794 &0.814 &0.822 &0.791 &0.766 &0.809 &0.802 &0.800 &0.783 &0.774\\
6024.3236 &0.937 &0.944 &0.966 &0.938 &0.943 &0.972 &0.943 &0.956 &0.949 &0.947 &0.944 &0.944 &0.935\\
6024.4521 &0.986 &0.992 &0.996 &0.986 &0.991 &1.014 &0.990 &0.995 &0.986 &1.004 &0.991 &1.032 &0.994\\
&&&&&&&&&&&\\ \hline
\end{tabular}
\end{table*}

\section{Comparison of calculated to observed velocities}

Table~4 reports calculated radial velocities and compares them to 
observations. O--C velocity for each echelle order and for each asteroid 
is plotted in Figure~1. Note that O-C differences are non-random. Each of the 
observing runs shows a distinctive trend. As already mentioned the telescope slit 
was kept along P.A.=90$^{o}$ (September 2004 observing run) and P.A.=270$^{o}$ 
(April 2005 run). Non-parallactic angle of the slit meant that, while the 
stellar image in red wavelengths was kept in the slit by the red-sensitive 
TV guider, the short wavelength image of the star was partly missing the slit. 
So asymmetric diffraction of the blue image on the spectrograph slit caused 
a shift in the angle of the beam entering the spectrograph and therefore an 
appreciable trend of radial velocity error with wavelength. We note that 
the two runs had the spectrograph mounted under opposite position angles, 
so Figure~1 shows opposite trends of velocity shifts. This interpretation is 
consistent with the fact that the dawn sky spectrum shows no trends of velocity 
error with wavelength. The reason is that in this case the slit was illuminated 
uniformly at all wavelengths, i.e.\ in the same way as for spectrum of a calibration 
lamp. As explained in Section~2 the final observed radial velocity of an asteroid 
was calculated without consideration of the bluest spectral orders, so this problem 
is largely avoided. Note that the observed velocities have errors which are still 
larger than the shot noise error of $\simeq 0.1$~km~s$^{-1}$. The reason are 
spectrograph flexures (Munari \&\ Lattanzi 1992) which could be accounted for only 
by more frequent calibration lamp exposures. 

Mean error in derived radial velocity is $\lsim 300$~m~s$^{-1}$. This is a much 
better value than previous results for this instrument (cf.\ Siviero et al.\ 2004). 
A similar analysis can be performed for any type of spectrograph. The possibility to 
study radial velocity error as a function of spectral order or position within the order
is superior to the usual velocity shift derived from telluric lines, which are 
almost exclusively confined to the red spectral region. Asteroid spectra are 
just reflected sunlight, so any unexpected change in spectral properties or 
radial velocity can be excluded. This is different from standard 
radial velocity stars for which velocity uncertainties are much larger.

\section{High resolution spectra of asteroids}

Low resolution spectra of asteroids depend on the variation of reflectance with 
wavelength. But the reflectance stays the same over the small width of spectral lines. Therefore 
normalized high resolution asteroid spectra are pure reflected sunlight with all the spectral lines of 
exactly the same shape and depth as in the Solar spectrum.  Table~5 \footnote{Available in electronic
form through CDS.} lists normalized observed spectra of asteroids and twilight together with a theoretical 
Solar spectrum (see Sec.~2). 
Presence of material intrinsic to the asteroid, 
such as regolith, manifests itself in low resolution spectra but is washed away by the 
normalization in the high resolution case.

This is demonstrated by Figure~2 where solid line plots the fraction of wavelength points with 
$dy = abs(I_1(\lambda)-I_2(\lambda))/I_2(\lambda)$ below a given value. Here $I$ denotes normalized 
intensity, and indices 1 and 2 pertain to mean asteroid and twilight spectrum (solid line), 
theoretical Kurucz and twilight spectrum (long dashes), or twilight and a constant of 1 (short 
dashes). One can read that a relative intensity difference of 3\%\ occurs in about 7\%\ of 
points if we compare asteroid and twilight spectrum, but this fraction increases to about 25\%\ 
if we compare observed twilight and theoretically computed spectra. Altogether some 35\%\ of all 
points have intensity at least 3\%\ fainter than the continuum level. There is no evidence for 
intrinsic differences between asteroid and twilight spectrum.

Observations of 
asteroids are a convenient way to assess the variation of resolving power between and along  
echelle orders. Solar spectrum has numerous absorption lines across the whole UV--IR domain. So it is 
better suited for this task than the sometimes scarce lines of the calibration lamp spectrum. 
Also, light from asteroids follows exactly the same path through atmosphere, telescope and the 
spectrograph as other objects on the sky. So it can be used to assess how the resolving power is 
influenced by atmosphere variations, accuracy of the telescope focus and of the centering of the 
object on the slit. Most importantly such observations can be used to study optical distortions 
and defocusing along the focal plane which causes a variation in resolving power along and between 
the spectral orders. An example of such variations is given in Figure~3  where a normalized 
theoretical Solar spectrum computed at a resolving power of 500,000 has been degraded until it provided 
the best match for the given 20~\AA\ wide part of the normalized observed spectrum. Variations are 
partly due to defocusing at the edges of spectral orders. On the other hand note that some spectral 
lines in the theoretical Kurucz spectrum do not match the observed twilight Solar spectrum, as manifested 
by the long-dashed line in Figure~2. The procedure therefore tried to compensate for mismatches by 
adjusting the resolving power, and this causes sharp jumps in Figure~3. 
A solution would be to replace a theoretical spectrum with a twilight or asteroid spectrum observed 
at a very high resolving power. Unfortunately no spectra of bright asteroids observed at 
$R\sim 100,000$ and with a high signal to noise ratio exist in the literature.

\section{Conclusions}

Observations of asteroids are convenient to measure radial velocity error as well as resolving power variation 
along and between the interference orders of an echelle spectrograph. They can be used to assess the 
accuracy of the absolute wavelength calibration on the level of m~s$^{-1}$. This is particularly valuable 
for wavelength calibration of slitless spectrographs, such as the radial velocity spectrograph 
on the forthcoming Gaia mission of ESA (Katz et al. 2004, Wilkinson et al. 2005). Rotation of 
asteroids with irregular shape and/or reflectance distribution over their surfaces can introduce radial 
velocity shifts on the m~s$^{-1}$ level. Most of 
the radial velocity standard stars are very bright and could be expected to have a certain degree of 
intrinsic spectrum variability. Asteroids, on the other hand, are observed at a wide range of magnitudes, 
so allowing a calibration suitable to the saturation limit of the telescope.
For the calibration of the GAIA radial velocity spectrometer (RVS), only
the brightest asteroids, i.e.\ those with a not too fancy shape, will be used.

Asteroid light follows the 
same path through the atmosphere, telescope and spectrograph as the light from other objects on the 
sky. The numerous absorption lines in the reflected Solar spectrum therefore allow a detailed mapping of 
line broadening due to variations in the resolving power along the spectrum. Solar spectrum is not 
expected to change much over time, moreover its light output is being constantly monitored. So it is 
safe to use the observed width of most individual Solar lines reflected from an asteroid to measure the 
resolving power of the spectrograph.  

Dedicated instruments are used to accurately map time variability in spectra of small areas on the Solar 
surface. But spectra reflected from asteroids offer a convenient way to measure the integrated 
high-resolution spectrum from the whole Solar disk using a normal telescope. So a variation of the high 
resolution integrated Solar spectrum with the Solar cycle could be measured.  

\acknowledgements{Generous allocation of observing time with the Asiago telescopes 
has been vital to this project. We would like to thank Fiorella Castelli, who 
continues to provide us with insights into the most recent versions of 
ATLAS and SYNTHE model atmosphere programs, and to Urtzi Jauregi for calculating the 
theoretical Kurucz spectrum. T.Z. acknowledges fruitful discussions 
with D.\ Hestroffer and hospitality of the GEPI group in Meudon. 
The financial support from the Slovenian Research Agency (to T.Z.) is kindly acknowledged.
We would like to thank the referee for very helpful and useful comments.}
\\

\end{document}